# Denoising convolutional neural networks for photoacoustic microscopy

Xianlin Song [a, #, *], Kanggao Tang [b, #], Jianshuang Wei [c, d, #], Lingfang Song [e]
[a] School of Information Engineering, Nanchang University, Nanchang 330031, China;
[b] Ji luan Academy, Nanchang University, Nanchang 330031, China;
[c] Britton Chance Center for Biomedical Photonics, Wuhan National Laboratory for Optoelectronics-Huazhong University of Science and Technology, Wuhan 430074, China;
[d] Moe Key Laboratory of Biomedical Photonics of Ministry of Education, Department of Biomedical Engineering, Huazhong University of Science and Technology, Wuhan 430074, China;
[e] Nanchang Normal University, Nanchang 330031, China;
# equally contributed to this work
* Corresponding author: songxianlin@ncu.edu.cn

## ABSTRACT

Photoacoustic imaging is a new imaging technology in recent years, which combines the advantages of high resolution and rich contrast of optical imaging with the advantages of high penetration depth of acoustic imaging. Photoacoustic imaging has been widely used in biomedical fields, such as brain imaging, tumor detection and so on. The signal-to-noise ratio (SNR) of image signals in photoacoustic imaging is generally low due to the limitation of laser pulse energy, electromagnetic interference in the external environment and system noise. In order to solve the problem of low SNR of photoacoustic images, we use feedforward denoising convolutional neural network to further process the obtained images, so as to obtain higher SNR images and improve image quality. We use Python language to manage the referenced Python external library through Anaconda, and build a feedforward noise-reducing convolutional neural network on Pycharm platform.We first processed and segmated a training set containing 400 images, and then used it for network training. Finally, we tested it with a series of cerebrovascular photoacoustic microscopy images.The results show that the peak signal-to-noise ratio (PSNR) of the image increases significantly before and after denoising.The experimental results verify that the feed-forward noise reduction convolutional neural network can effectively improve the quality of photoacoustic microscopic images, which provides a good foundation for the subsequent biomedical research.

## 1. INTRODUCTION

Photoacoustic imaging is a biomedical imaging mode developed rapidly in recent years, which has the advantages of high contrast of optical imaging and high resolution of ultrasonic imaging in deep biological tissues. Photoacoustic Microscopy (PAM) is an important way to achieve Photoacoustic imaging[1]. According to the size of the focal spots of beam and acoustic beam in the imaging system, PAM can be further distinguished as OR-PAM and AR-PAM[7]. Regardless of the system, due to the extremely low conversion efficiency of photoacoustic signals, the resulting image has a particularly low Peak Signal to Noise Ratio (PSNR), especially in the case of weak laser power and deep penetration, where useful information is almost drowned in the noise[1]. At present, there are many methods to de-noising photoacoustic images, such as the most common median filtering de-noising and wavelet domain de-noising. In the computer science field, the famous denoising algorithms are the BM3D (Block-Matching and 3D Filtering) algorithm, and other algorithms similar to the PGPD signal denoising algorithm, which takes a long time and has low efficiency. These algorithms and their derived optimized algorithms have been widely used, but the usage of Denoising Convolutional Neural Network (DnCNN) for photoacoustic microscopy image denoising is still rare. For this reason, we try to study a feasible neural network to improve the PSNR of images on the premise of ensuring the Structural Similarity Index (SSIM).

Denoising Convolutional Neural Network is a new deep learning neural network which is greatly enhanced on the basis of the traditional denoising neural network and innovatively realizes the learning for the residual image by drawing on the concept of Residual learning from Residual Network (ResNet)[6]. It also adds Batch Normalization (BN), which helps the neural network obtain good training results even when the convolutional layers of neural networks are very deep. In this experiment, the convolutional neural network we use consists of three different types of layers. The first layer is Conv2D+ReLU, using 64 3*3* C filters to generate 64 feature graphs. Here c equals to 1, which means that the image we use is black-and-white image. Then we add 20 layers of Conv and ReLU to the neural network. In the DnCNN, the sensing field with layer depth d should be (2d +1) * (2d +1)[6]. Increasing the size of the sensing field can make use of more graphic information, but it also brings lower efficiency. Considering the size of the training images used (256*256 pixels), we set the depth to 20 and added batch normalization to each layer, so that each layer should actually be Conv2D+ReLU+BN.The size of the filters are 3*3*64.The last layer is the Conv layer, and c (actually 1) filters at the size of 3*3*64 are used to reconstruct the output image. At the same time, the residual image can be learned and predicted by a residual unit to improve the efficiency of the neural network. The entire structure of the neural network is demonstrated in figure 1.

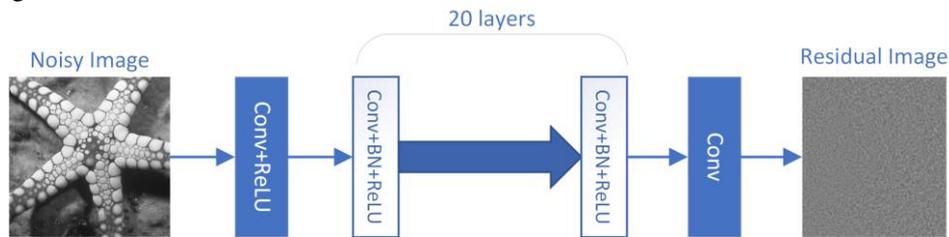

Figure 1. The structure of the DnCNN we set and the final output image. The number of hidden layers set in this experiment is 20.

## 2. METHODS

**2.1 Residual learning and batch standardization in the DnCNN**

One of the biggest characteristics of DnCNN is to learn from the Residual Network(ResNet) and uses residual learning[6]. The residual network was first proposed by 4 scholars of Microsoft Research. The background is that, during the construction of the traditional neural network, when the depth of the neural network reaches a certain level, it may cause the problem that is called the Exploding Gradient or the Vanishing Gradient[2]. Although the neural network with greater depth can be converged by the way of specification initialization and the introduction of median specification layer, with the increase of the depth of the network, the accuracy of pattern recognition still begins to decline after reaching the saturation. This situation fully shows that the traditional neural network cannot be simply optimized. The residual originally refers to the deviation between the measured value and the predicted value, which means that the learning of the residual can be generally understood as the learning of the deviation. The authors of residual network hope to fit a residual mapping with multi-layer network instead of only a few layers, so a multi-layer neural network is formed for the application of image recognition[4]. DnCNN did not completely copy the residual learning part of the residual network. Different from the multi-layer network-type residual unit in the residual network, DnCNN only relies on a single residual unit to predict the residual image. Figure 2 shows the residual learning part in the residual neural network and the DnCNN.

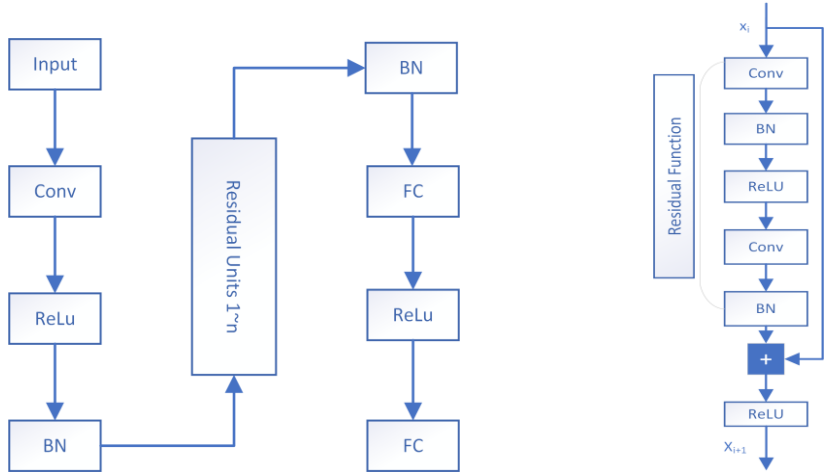

Figure 2. The left figure is the framework of the residual network, which is an important part of the deep residual network. The figure on the right side shows a single residual unit. The residual network framework is composed of multiple residual units, and only one residual unit, as shown in the right figure, is used in DnCNN to predict the results.

Suppose the input of the DnCNN is regarded as y=x + v, where x is the clean image and V is the noise contained in the image. For the ordinary denoising networks, the primary purpose is to predict possible denoised images through network learning, that is, to predict F(y)=x through network learning. However, in the DnCNN, we use residual learning to train the neural network, so x will be indirectly obtained, namely x= y-H (y), where H(y) is the expected learning function of the neural network. The loss function of the neural network is the mean square error between the predicted residual image and the real residual image, and its calculation formula[6] is as follows:

$$l(\theta) = \frac{1}{2N} \sum_{i=1}^{N} \| R(y_i - \theta) - (y_i - X_i) \|_F^2 \quad (1)$$

In practical situation, the relationship between clean image and noise is not simple addition and subtraction, but the basic theory of learning and prediction for residual remains unchanged.

The second major feature of the DnCNN is the usage of Batch Normalization(BN)[6]. The BN algorithm is also designed to solve the problem when the number of layers of neural network deepens. In the neural network, the input of each layer will inevitably make the initial corresponding input signal distribution different after the operation in this layer, and the distribution deviation of the shallow layer in the network will gradually be amplified by the deep layer[8]. When the network layers are relatively deep, this extremely uneven distribution will seriously affect the convergence rate of the network[5]. The BN algorithm can be used to normalize the input of some or all layers to keep them in a stable distribution (for example, a normal distribution). Even when the depth reaches dozens or even hundreds of layers, the convergence speed of the network will not decline significantly[6].

## 2.2 Image processing and network training

Before the neural network is trained, the training images are preprocessed. First of all, we uniformly grayscale them. On the one hand, our network only supports single-channel images at present. On the other hand, in the practical application process, the photoacoustic microscope images have better observation effect when the contrast color is strong. After that, use the resize function in the openCv library to convert the image to a uniform size of 256*256 with the bit depth of 8. As we have only 400 images in our training set, we have put in data augmentation to the dataset. Since an untrained neural network determines two images that have undergone only minor changes to be different, we can use this method to increase the size of the training set. The specific operation is to rotate each picture in the dataset by 90 °/180 °/270 °, and then a dataset with an actual capacity of 1600 pictures can be obtained, which is sufficient for the training of neural network. After that, we will repair these rotated pictures and save the training set files into the npy folder for later use.

The npy folder is saved because these types of files are faster when read, making it easier to train the entire network. For the trained network, we used a test data set of 68 natural images from the Berkeley Split Dataset (BSD68)[6], and another test set of 12 images. During the test, we first measured the PSNR value of the original image, and then tested the PSNR and SSIM value of each image after the operation, and saved these two data in the table.

In the course of network training, we found that each training process takes quite a long time. On the one hand, due to the graphics card configuration problem, the program can only support CPU computing, but not collaborative computing by GPU. On the other hand, the training speed is affected by the training times on the same dataset. In the initial design, we plan to train the network on the training set for 50 times, and at the same time monitor the change of loss function value. In the first three training sessions, the value of loss function changes very significantly. In the 3rd to 10th training sessions, the average value of loss function can be reduced by 0.001 per two training sessions. The figure 3 shows the significant change in the value of loss function during the first training session.

```
Epoch 1/10
    7/1862 [..........................] - ETA: 3:20:16 - loss: 0.0189
Epoch 1/10
  540/1862 [=======>..................] - ETA: 2:32:11 - loss: 0.0024
```

Figure 3. It can be clearly seen that the value of loss function has decreased significantly before the end of the first training session.

With the increase of training times, the change of loss function becomes less and less obvious, and even approaches saturation when trained for 20 times. In this case, we improved the program and saved the training model every 5 times so that we can control the variables and test each model. Then we used these models to perform denoising analysis on the same image and recorded the values of PSNR and SSIM. The test results show that concerning both efficiency and effect, 10 times of training can achieve a more ideal effect under the premise of high efficiency.

## 3. RESULTS

The image of zebrafish obtained by photoacoustic microscopy was used as a test image, and the model trained for ten times and trained for five times was used as a test model to conduct noise addition/denoising test respectively. Meanwhile, in order to compare this algorithm with the BM3D algorithm which is relatively better in the traditional denoising algorithms, we used the BM3D algorithm written in C++ language to test the same image (the platform was Visual Studio 2019). Respectively record the PSNR before and after the denoising by these algorithms. The results are as follows.

Table 1: PSNR after noise addition/denoising under three different conditions

| Algorithm \ Situation | After add noise | After denoising |
|---|---|---|
| DnCNN_10 | 20.1252 | 31.2829 |
| DnCNN_4 | 20.1705 | 30.3901 |
| BM3D | 20.2233 | 29.8493 |

It can be seen from the data above that, from the perspective of PSNR alone, the DnCNN trained for 10 times is superior to DnCNN trained for 4 times and the BM3D algorithm. At the same time, we also measured the SSIM value of images processed by the DnCNN, which was 0.9117 and 0.8916 respectively under 10 and 4 training conditions. This also indicates that DnCNN trained for 10 times can retain the information of the original image to a greater extent. In consideration of the fact that photoacoustic microscopy generally does not use colorless gray-scale images in practical applications, we converted the processed gray-scale images into hot images through Matlab. The images are shown as figure 4.

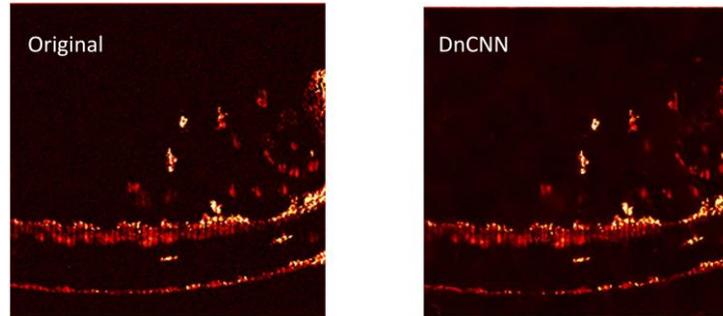

Figure 4. the left image is part of photoacoustic imaging of zebrafish used for testing; the right image is after denoising. It is obvious that the noise in the background of the left figure is removed, and the tissue noise near the blood vessels is also removed. The image after denoising retains the basic detail structure of the original image and achieves good denoising effect.

When comparing DnCNN with the traditional denoising algorithm, we used the unprocessed grayscale image instead of the hot image, because the grayscale image can show the difference between the two more intuitively than the hot image. As is shown below, image denoised by BM3D algorithm is more blurred, more fine structures are lost, and even some important structures are blurred. This is unacceptable for biomedical applications that require precision. In terms of running time, the processing time of BM3D algorithm is 118.53s, while DnCNN only uses 2.34s. The results are shown as figure 5.

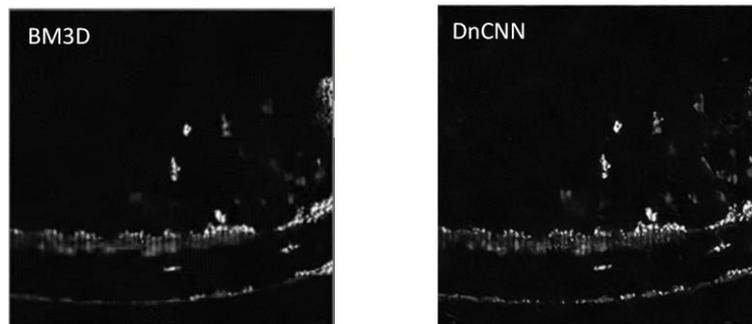

Figure 5. The left image is the one processed by BM3D algorithm; the right image is the one processed by DnCNN. It is obvious from the diagram that the right image is clearer and retains more detaisl than the left one.This may be due to the difference in image processing between the two denoising methods.

## 4. CONCLUSIONS

In the field of photoacoustic image denoising, DnCNN has its unique advantages. Residual learning can help DnCNN save a lot of time and is more accurate than traditional denoising methods. Batch normalization enables the network to maintain a stable convergence rate at a great depth, which further improves the efficiency of the neural network. The size of the training set and the number of training times on the same training set can significantly affect the training result of DnCNN. The larger the training set is and the more training times it runs, the higher accuracy the neural network can obtain. Compared with the traditional BM3D algorithm, DnCNN with sufficient training times has better denoising effect and shorter denoising time, and can fully retain the details of the original image, so it can better meet the requirements of accuracy in the biomedical field. To sum up, DnCNN with sufficient training times is more suitable for photoacoustic imaging denoising neural network.

However, there is still room for improvement. On the one hand, since the Nividia GTX 1650 graphics card used in this experiment does not support GPU computing, only CPU is used for computing in the process of program running, but the program itself can support GPU computing. In other words, under the condition of proper configuration, GPU computing will bring faster training speed and image processing speed to the program. In this way, more hidden layers of Conv+BN+ReLU can be added to the network, and the optimal equilibrium point of training efficiency and training effect can be improved, and the image denoising effect can also be further improved.